\documentclass[
reprint,
superscriptaddress,
amsmath,amssymb,
aps,
]{revtex4-1}

\usepackage{graphicx,epsf}

\begin{document}

\title{Gap prediction in hybrid graphene - hexagonal boron nitride nanoflakes\\ using artificial neural networks}

\author{G. A. Nemnes}
\email{nemnes@solid.fizica.unibuc.ro}
\affiliation{University of Bucharest, Faculty of Physics, Materials and Devices for Electronics and Optoelectronics Research Center, 077125 Magurele-Ilfov, Romania}
\affiliation{Horia Hulubei National Institute for Physics and Nuclear Engineering, 077126 Magurele-Ilfov, Romania}
\author{T. L. Mitran}
\affiliation{Horia Hulubei National Institute for Physics and Nuclear Engineering, 077126 Magurele-Ilfov, Romania}
\author{A. Manolescu}
\affiliation{School of Science and Engineering, Reykjavik University, Menntavegur 1, IS-101 Reykjavik, Iceland}

\begin{abstract}
The electronic properties graphene nanoflakes (GNFs) with embedded hexagonal boron nitride (hBN) domains are investigated by combined {\it ab initio} density functional theory calculations and machine learning techniques. The energy gaps of the quasi-0D graphene based systems, defined as the differences between LUMO and HOMO energies, depend on the sizes of the hBN domains relative to the size of the pristine graphene nanoflake, but also on the position of the hBN domain. The range of the energy gaps for different configurations is increasing as the hBN domains get larger. We develop two artificial neural network (ANN) models able to reproduce the gap energies with high accuracies and investigate the tunability of the energy gap, by considering a set of GNFs with embedded rectangular hBN domains. In one ANN model, the input is in one-to-one correspondence with the atoms in the GNF, while in the second model the inputs account for basic structures in the GNF, allowing potential use in up-scaled structures. We perform a statistical analysis over different configurations of ANNs to optimize the network structure. The trained ANNs provide a correlation between the atomic system configuration and the magnitude of the energy gaps, which may be regarded as an efficient tool for optimizing the design of nanostructured graphene based materials for specific electronic properties.
\end{abstract}

\maketitle

\section{Introduction}

The absence of an electronic gap in pristine graphene hinders many of the expected applications based on the field effect. Graphene nanopatterning is one way to tune the electronic and transport properties and this can be achieved by reducing the dimensionality \cite{PhysRevLett.97.216803,Son2006,doi:10.1021/nl0617033,C0JM00261E}, by drilling periodic arrangements of holes \cite{Bai2010,C7TC00029D}, by embedding hexagonal boron-nitride (hBN) \cite{Ci2010,1882-0786-5-8-085102,C2NR32201C,NEMNES1,NEMNES20131347,Chen2017} or a combination of these. Graphene nanoribbons (GNRs) and graphene nanoflakes (GNFs), typically passivated with monovalent species like hydrogen or halogen atoms, are two examples of quasi-1D and quasi-0D graphene systems, respectively, which attracted a lot of attention in the past few years. GNRs can have a metallic or semiconducting behavior depending on the lateral width and edge type, armchair or zigzag. In contrast to GNRs, where only the edge states may influence the electronic properties, in GNFs these are markedly influenced by both edge and corner states and, in general, by the different possible shapes \cite{snook,doi:10.1063/1.4953172}. In addition, GNFs may be functionalized, which further extends the range of the electronic, optical and magnetic properties.

GNFs can be produced by bottom-up approaches, where the synthesis takes place in solution by mechanical extrusion, using magnetic field alignment and thermal annealing \cite{doi:10.1021/cr068010r,B717585J} or by top-down methods, using techniques like e-beam lithography \cite{Berger1191}, plasma etching \cite{doi:10.1002/smll.201000291},
or a cationic surfactant mediated exfoliation of graphite \cite{MUTYALA2015692}. 
Besides the many applications envisioned for nanoelectronics and spintronics \cite{doi:10.1021/acs.nanolett.8b00453}, 
more recently, novel applications also indicate the role of GNFs for biological recognition \cite{Castagnola2018}. Therefore, methods for an efficient investigation of multiple configurations of GNFs and related structures are highly demanded.

In the past few years, machine-learning (ML) techniques are gaining ground in the field of condensed matter. They have been developed to predict the band gaps in solids \cite{PhysRevB.93.115104,PILANIA2017156}, while they also provide new clues in crystal structure prediction \cite{LIU2017159,doi:10.1021/jacs.8b03913}. They can be used to bypass the Kohn-Sham equations by learning energy functionals via examples \cite{Brockherde2017} or predicting DFT Hamiltonians \cite{Hegde2017}. The generic aim is to develop less expensive and faster methods to calculate the system's properties. To this end, the methodology contained in PROPhet \cite{Kolb2017} provides a general framework for coupling machine learning and first-principles methods.
ML techniques can also provide more insights about the physical properties of a system. The usefulness as a universal descriptor of grain boundary systems was pointed out \cite{Rosenbrock2017}, potentially indicating which building blocks map to particular physical properties.
ML techniques can also achieve high accuracies, the prediction errors of molecular machine learning models being below that of the hybrid DFT error \cite{doi:10.1021/acs.jctc.7b00577}.

Regarding graphene systems, ML techniques have been employed in several studies, e.g. for obtaining an accurate interatomic potential for graphene \cite{PhysRevB.97.054303}, searching the most stable structures of doped boron atoms in graphene \cite{doi:10.1063/1.5018065}, for investigating the influence of GNF topology \cite{doi:10.1021/acscombsci.6b00094} and predicting accuracy differences between different levels of theory \cite{0957-4484-28-38-38LT03}, as well as for the prediction of interfacial thermal resistance between graphene and hBN \cite{C8NR05703F}.

In this paper we investigate the electronic properties of hybrid graphene - hBN nanoflakes, using combined DFT and ML methods. We construct the distribution of gap energies using {\it ab initio} DFT calculations, as LUMO-HOMO differences, which depend on the size and position of the hBN domains within the GNF. Given the large number of possibilities of setting the hBN domains, extensive DFT calculations are typically required, with a significant computational cost. Instead, we develop artificial neural network (ANN) models able to reproduce the energy gaps with high accuracies, which significantly reduce the computational effort. We test our ANN models against reference gap values obtained by DFT and discuss the optimal conditions for the network structure.

\section{Model systems and computational methods}

\begin{figure}[t]
\centering
\includegraphics[width=7.cm]{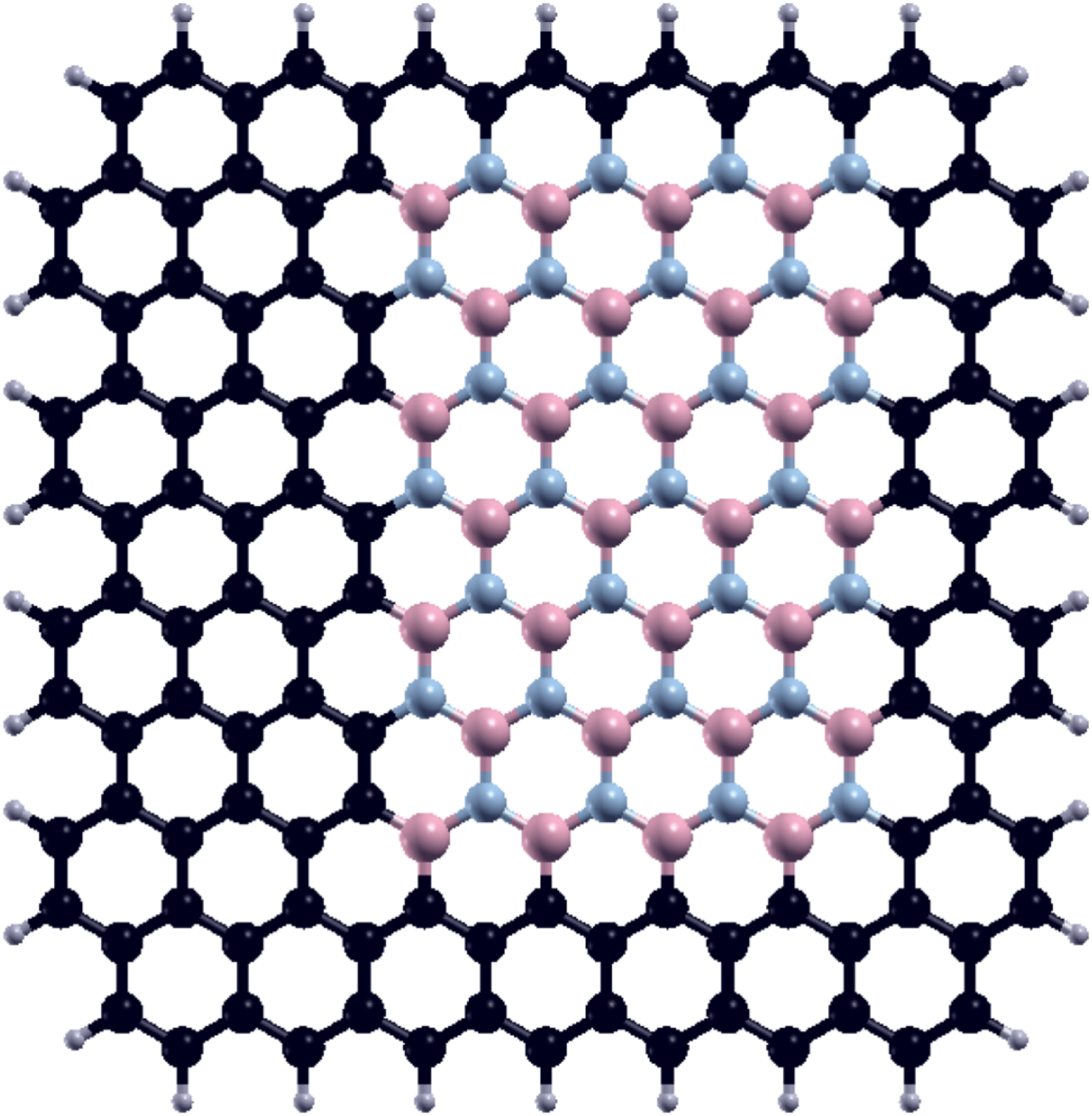}
\caption{A typical graphene nanoflake with an embedded rectangular hBN domain. The edges are passivated with hydrogen. Each system contains $N = 166$ C, B, or N atoms, colored in black, ping and light blue respectively, and $N_{\rm H} = 34$ hydrogen atoms. 
}
\label{gnf}
\end{figure}

We consider GNFs with embedded hBN domains, passivated with hydrogen, as indicated in Fig.\ \ref{gnf}. The hBN domains are rectangular shaped regions containing an equal number of boron and nitrogen atoms. In this way the systems retain an intrinsic semiconducting behavior, without a net chemical doping. The rectangular hBN embedding is randomly positioned in the graphene nanoflake. The widths and heights of the rectangular hBN regions are extracted from a flat distribution so that the entire graphene nanoflake can be replaced by BN. The systems analyzed here have a total of 200 atoms, of which $N = 166$ atoms are stemming from graphene/hBN and $N_{\rm H} = 34$ hydrogen atoms. For the investigation of the electronic properties, a number of 900 non-equivalent systems are generated.

The DFT calculations are performed using the SIESTA code \cite{0953-8984-14-11-302} employing local density approximation (LDA) in the parametrization of Ceperley and Alder \cite{PhysRevLett.45.566}. The strictly localized basis set allows a linear scaling the computational time with the system size. The self-consistent solution of Kohn-Sham equations was obtained using the standard double-$\zeta$ polarized basis set, a grid cutoff of 100 Ry and norm-conserving pseudopotentials of Troullier and Martins \cite{PhysRevB.43.1993} with a typical valence electron configurations for carbon, boron and nitrogen. The gap energies are determined, being defined as the difference between the LUMO and HOMO energies, $E_{\rm gap} = E_{\rm LUMO} - E_{\rm HOMO}$. 

Based on the DFT results we implement ANN models able to reproduce the gap energy for similar systems from a new set. The ANNs are standard fully connected backpropagation neural networks implemented using the FANN library \cite{nissen03}, with three layers: one input layer, one hidden layer and one output layer. In {\it Method 1} we assign an input neuron to each atom of species C, B or N, so that the number of input neurons is $N_{\rm in} = 166$. {\it Method 2} accounts for the chemical neighborhood of a certain atomic species and its prevalence in the system. In this case, we use $N_{\rm in} = 20$ input neurons, where 4 of them account for the proportions of the four atomic species (C, B, N, H) and 16 neurons are associated with the normalized counts of $(X_i, Y_1, Y_2, Y_3)$ atom quadruplets, where $X_i =$ C, B, N and $Y_{1,2,3}$ are the three nearest neighbors of $X_i$, with $1 \le i \le N$. The number of neurons in the hidden layer $N_{\rm h}$ is varied, from 25 to 200 neurons, in order to find a close-to optimal configuration. The output layer has a single neuron, $N_{\rm out} = 1$, and the result maps the gap energy by a continuous function in the $[0,1]$ interval, corresponding to a maximum gap energy $E_{\rm gap}^{\rm max} = 4$ eV. {\it Method 2} has the advantage that the input does not depend on the system size, allowing the same ANN to handle up-scaled structures.

For training we employ the iRPROP algorithm of Igel and H\"usken \cite{Igel00}, which is a variant of the standard RPROP algorithm introduced by Riedmiller and Braun \cite{298623}. The iRPROP algorithm is adaptative and there is no pre-set learning rate. The sigmoid activation function is used and the mean square error during training is set to $10^{-5}$. Since the ANNs are randomly initialized and the final weight configurations depend on the seeds, an ensemble of 1000 ANNs is trained on the same data set. Finally a statistics regarding the accuracy obtained on the test data is performed.

The trained ANNs are tested on a set of 100 new examples and the predicted gaps are compared to the reference values obtained by DFT calculations. We use the R$^2$ coefficient of determination as a measure of how far the observed outcomes are replicated by the ANN model.

\section{Results and discussion}

GNFs are quasi-0D systems with a discrete energy spectrum, where the gap energy is typically influenced by their geometry, passivation and nanopatterning. By embedding hBN in GNFs, which is a wide band gap isomorph of graphene, it is expected that the gap energy has a strong variation. Particularly in finite systems, the position and shape of the embedded rectangular hBN domain, closer to the edges or at the GNF center, significantly influences $E_{\rm gap}$.   

\begin{figure}[t]
\centering
\includegraphics[width=\columnwidth]{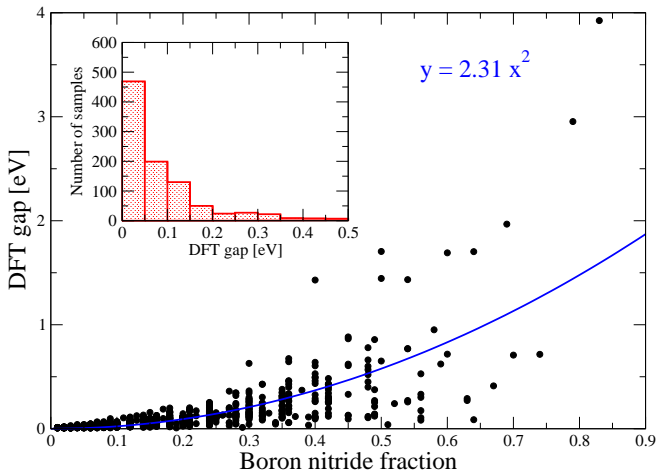}
\caption{The reference DFT gap vs. the BN fraction $f_{\rm BN}$. Depending on the position and shape of the BN rectangular domain, different gap values are obtained at the same $f_{\rm BN}$. A fit with a second degree polynomial function shows the statistical increase of $E_{\rm gap}$ with $f_{\rm BN}$ ($y=2.31x^2$). The inset shows a histogram of the DFT gap values, focusing on the small energy gaps.}
\label{BNconc}
\end{figure}

We first investigate the variation of $E_{\rm gap}$ as a function of the hBN domain size, given by the BN fraction $f_{\rm BN} = (N_{\rm B} + N_{\rm N}) / N$, where $N_{\rm B}$ and $N_{\rm N}$ are the number of boron and nitrogen atoms, respectively. As it is indicated in Fig.\ \ref{BNconc}, there a rather wide dispersion of values, as there are multiple configurations with the same $f_{\rm BN}$. Still, a clear trend is visible for $E_{\rm gap}(f_{\rm BN})$: larger gaps may be obtained as the BN domain size increases, while smaller gaps are still present. A fit with a second degree polynomial function shows the statistical increase of the gap energy, as $E_{\rm gap} = 2.31 f^2_{\rm BN}$. 

\begin{figure}[t]
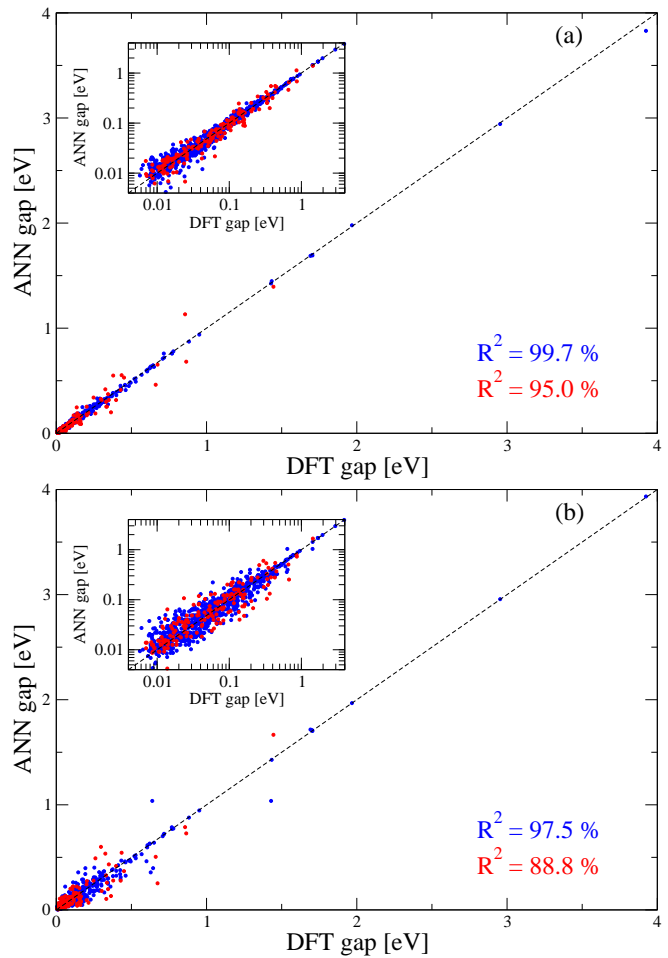

\centering
\includegraphics[width=\columnwidth]{figure3a}
\includegraphics[width=\columnwidth]{figure3b}
\caption{Predicted ANN gap vs. reference DFT gap, for typical fully connected networks with three layers: (a) {\it Method 1} (166/100/1 neurons) and (b) {\it Method 2} (20/100/1 neurons). The results corresponding to the training and test sets are represented in blue and red colors, respectively. The $R^2$ coefficient of determination is calculated for both training and test examples. The inset contains $\log-\log$ plots showing a detail view over the small gap energy range.}
\label{ann100}
\end{figure}

Next, we investigate the accuracies in predicting the energy gaps for the proposed ANN models. In {\it Method 1} we start with an ANN configuration with three layers, with $N_{\rm in} = 166$ neurons in the input layer, $N_{\rm h} = 100$ neurons in the hidden layer and $N_{\rm out} = 1$ output neuron. The ANN is trained on 800 examples and tested on a new set of 100 structures. The results are represented in Fig.\ \ref{ann100}, where the predicted gap is plotted vs. the reference DFT gap. 
The $R^2$ coefficient of determination calculated for the training set yields a rather high value of $99.7\%$, which indicates a consistent convergence during training. Typically, for this ANN configuration, the threshold for mean square error set to $10^{-5}$ is reached in $\sim 400$ steps. Running the ANN on the test systems, one obtains $R^2$ values as high as $95\%$. However, as detailed in the following, the performance of the ANN relies on the converged configuration, which may depend on the ANN initialization.     

In the second method, labeled {\it Method 2}, the ANN is trained to capture the local chemical neighborhood. For the same set of systems, there are 16 instances of atom quadruplets $(X,Y_1,Y_2,Y_3)$, with $X =$ C, B, N and $Y_{1,2,3} =$ C, B, N, H. These are counted for each structure and normalized to $N$, the total number of carbon, boron and nitrogen atoms. Along with these 16 inputs, the fractions corresponding to each of the four atomic species are added, yielding a total number of $N_{\rm in} = 20$ input neurons. These extra input neurons improve the prediction behavior of the ANN as they emphasize the importance of the size of the hBN domains. The same training procedure and convergence criterion are employed as for {\it Method 1}. The convergence during training is poorer ($R^2 =97.5\%$) and the obtained accuracy is typically smaller ($R^2=88.8\%$) for {\it Method 2}, although comparable with the ones obtained for {\it Method 1}. However, {\it Method 2} is by construction scale invariant and this is potentially a significant advantage in investigating systems with different sizes.

\begin{figure}[t]
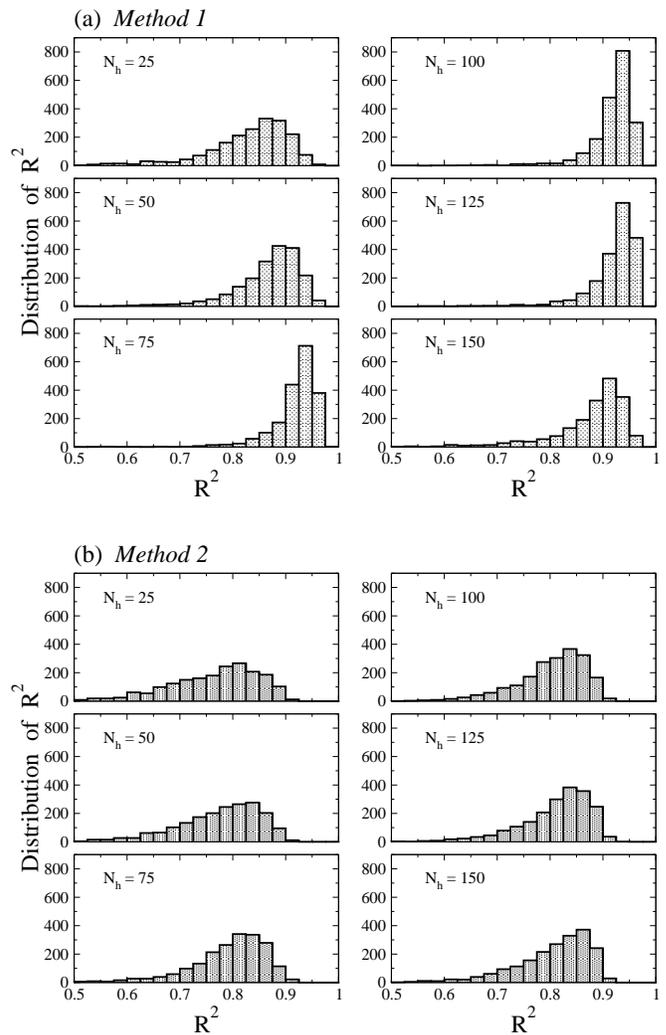

\centering
\includegraphics[width=\columnwidth]{figure4a} \vspace*{0.2cm}\\
\includegraphics[width=\columnwidth]{figure4b}
\caption{Histograms of $R^2$ coefficients for (a) {\it Method 1} and (b) {\it Method 2} corresponding to different number of neurons in the hidden layer $N_{\rm h}$. An ensemble of 2000 starting configurations of ANNs were considered to construct the $R^2$ distributions.}
\label{R2stat-1}
\end{figure}

The final ANN configuration following the training phase depends on the assigned random initial weights. Consequently, the accuracy of the output results obtained by running the test examples is subject to the initialization procedure. In order to see how robust are the obtained results, we construct histograms using an ensemble of 2000 trained ANNs. The results are shown in Fig.\ \ref{R2stat-1} for the two methods. In {\it Method 1}, as the number of hidden neurons is varied, the distributions evolve from a rather wide-spread distribution of $R^2$ coefficients for $N_{\rm h} = 25$ to a more confined distribution around the high accuracy values, for a number of neurons in the hidden layer, $N_h$, between 100 and 125 neurons. Increasing further $N_h$ does not improve the accuracy. Rather, as the ANN becomes larger, memory effects become important to the detriment of capturing the essential features of the structures. Moreover, by decreasing the mean square error to $10^{-6}$ when training ANNs with $N_{\rm h} > 150$, they become over-trained and the $R^2$ coefficient does not improve either. Therefore, we conclude that optimal ANN configurations exist, with quite high maximal output accuracies ($\sim 97$ \%) and a relatively narrow band of $\sim 10$ \%, where the $R^2$ coefficients of the most trained ANNs can be found.

\begin{figure}[h]
\centering
\includegraphics[width = 8 cm]{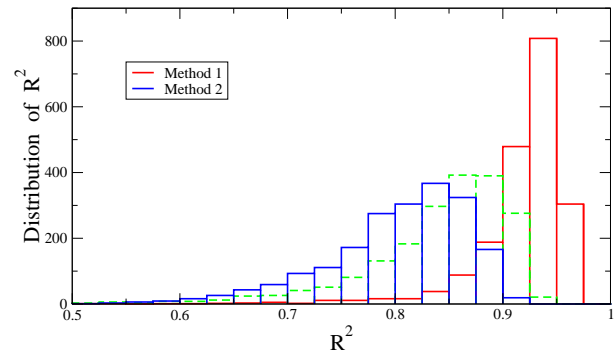}
\caption{Histograms of $R^2$ coefficients, comparing {\it Method 1} (red) and {\it Method 2} (blue) for $N_{\rm h} = 100$ neurons. Additionally, ANNs trained only on the geometrical properties of the hBN domains (green) perform statistically better than {\it Method 2}, but worse than {\it Method 1}.}
\label{R2stat-1-2-3}
\end{figure}

Comparatively, employing {\it Method 2}, the $R^2$ histograms follow the same trend, although the accuracy spread is larger. Still, the highest values can reach as high as $\sim 91\%$. This shows that by describing the local chemical environment and constructing a statistics reflecting the neighborhood of the different species, one can infer quite reasonably the electronic features of the GNFs, in particular the energy gaps. A direct comparison to {\it Method 1} is shown in Fig.\ \ref{R2stat-1-2-3}. Additionally, the distribution of $R^2$ coefficients for an intermediate model based on geometrical parameters of the rectangular hBN domains is also indicated. In this case, the four distances between the edges of the hBN rectangles and the edges of the GNF along with the two linear sizes of the hBN domains were taken as inputs, i.e. $N_{\rm in} = 6 $ input neurons. However, this approach can be used as long as the geometric features of the samples can be easily identified, i.e. in this case the parameters describing the rectangular shapes. The distribution of $R^2$ coefficients lies in between the ones corresponding to {\it Method 1} and {\it Method 2}, with a maximum at 94.1\%, compared to the best results of 97.2\% obtained with {\it Method 1} and 91.9\% using {\it Method 2}. This also shows that by identifying the geometrical features in graphene-hBN systems, without taking into account a detailed representation of the species present in the structure, i.e. considering the hBN domain as a whole, reasonable accuracies may be achieved.

\clearpage

\section{Conclusions}

The electronic properties of GNFs with embedded hBN domains were investigated using combined DFT and ML techniques. Using DFT calculations we constructed the energy gap distribution for a set of systems with different rectangular hBN shapes. The collected data was used to train two types of ANNs. In {\it Method 1}, one input neuron is assigned to one atom of species C, B or N, while in {\it Method 2} the prevalence of the chemical neighborhood and atomic species was taken into account.
The trained ANNs provide a correlation between the different domain shapes, their sizes and location within the GNFs, on one hand, and the magnitude of the energy gaps, on the other hand. 
{\it Method 1} shows the highest accuracies, while in {\it Method 2} smaller ANNs are not bound to a fixed system size and the accuracies are comparable. A statistical analysis reveals the optimal configurations of the three layer ANNs, pointing out potential memory and over-training effects in large networks.
The approach based on ANNs is therefore a feasible route, providing a reduction of the computational effort, while retaining a high accuracy and therefore may be employed for optimizing the design and selecting candidates of nanostructured graphene based materials for specific electronic properties. \\

{\bf Acknowledgments} \\

This work was supported by the Romanian Ministry of Research and Innovation under the project PN 18090205/2018 and by Romania-JINR cooperation project JINR Order 322/21.05.2018, no. 29.


\vspace*{-0.2cm}

\section*{References}

\bibliographystyle{elsarticle-num}
\bibliography{manuscript}

\end{document}